\documentstyle[epsf,twocolumn,seceq]{jpsj}

\def\runtitle{Filling control of the Mott insulator Ca$_{2}$RuO$_{4}$}
\def\runauthor{Hideto {\sc Fukazawa} and Yoshiteru {\sc Maeno}}

\title
{
Filling Control of the Mott Insulator Ca$_{2}$RuO$_{4}$
}

\author
{ 
Hideto {\sc Fukazawa}$^{1,}$\footnote{E-mail address: hideto@scphys.kyoto-u.ac.jp}
and Yoshiteru {\sc Maeno}$^{1,2}$
}

\inst
{
$^1$Department of Physics, Kyoto University, Kyoto 606-8502\\
$^2$Core Research for Evolutional Science and Technology, 
Japan Science and Technology Corporation\\
(CREST-JST), Kawaguchi 332-0012
}

\recdate
{
\today
}

\abst
{
We have grown single crystals of electron doping system 
Ca$_{2-\it{x}}$La$_{\it{x}}$RuO$_{4}$ (0.00 $\leq$ $x$ $\leq$ 0.20) 
by a floating zone method. 
The first order metal/non-metal transition and canted antiferromagnetic 
ordering occur for $0.00 < x < 0.15$, similar to those in the bandwidth 
controlled system Ca$_{2-\it{x}}$Sr$_{\it{x}}$RuO$_{4}$ (CSRO). 
However, comparing with CSRO, we found a rather different 
metallic ground state adjacent to the non-metallic ground state 
with canted antiferromagnetic order. 
Instead of short-range antiferromagnetic correlation found in CSRO 
($0.20 \leq x < 0.50$), the metallic ground state of the present system 
is characterized by strong ferromagnetic correlation. 
}

\kword
{
Ca$_{2}$RuO$_{4}$, Mott insulator, metal-insulator transition, 
ruthenate, filling control, Sr$_{2}$RuO$_{4}$
Mott transition, floating zone method
}

\begin{document}
\sloppy
\maketitle

\section{Introduction}
The discovery of superconductivity in Sr$_{2}$RuO$_{4}$ 
($T_{\rm{c}} \simeq 1.5\; \rm{K}$)
has attracted strong research interests 
because it is the first layered-perovskite 
superconductor without copper.~\cite{mae1} 
In addition its spin-triplet superconductivity has promoted deeper 
investigation of its anisotropic and unconventional physical properties in 
both superconducting and normal states.~\cite{ish1,mac1,mao1,nis1,mae2,muk1} 
The electronic properties of Sr$_{2}$RuO$_{4}$ reflecting 
strong electronic correlation imply 
that it is in the vicinity of Mott transition.~\cite{mae1} 
In fact its isomorph Ca$_{2}$RuO$_{4}$ is a Mott insulator, 
which exhibits canted antiferromagnetic (CAF) ordering at 113 K 
(= $\it{T}_{\rm{N}}$: N\'{e}el temperature) and 
insulator-"near" metal transition at 357 K (= $\it{T}_{\rm{MI}}$) 
on heating.~\cite{nak1,cao1,bra1,ale1,fuk1} 

Recently it has been recognized that orbital degrees of freedom of 
$\it{t}_{\rm{2g}}$ states, originating from the low-spin $4d^{4}$ 
configuration of Ru$^{4+}$ ions, play important roles 
in physical properties of quasi-two-dimensional Mott-transition system 
Ca$_{2-\it{x}}$Sr$_{\it{x}}$RuO$_{4}$ (CSRO).~\cite{nak2,nak3} 
In CSRO, the metal-insulator (MI) or strictly speaking, 
metal/non-metal (M/NM) transition, occurs simultaneously 
with the first-order structural phase transition, 
which is attributable to the Jahn-Teller effect.~\cite{nak3,fri1} 
Smaller ionic radius of Ca$^{2+}$ ions ($r_{\rm{Ca}}=1.18\; \rm{\AA}$) 
compared with that of Sr$^{2+}$ ions ($r_{\rm{Sr}}=1.31\; \rm{\AA}$) 
brings about the transition from quasi-tetragonal orthorhombic L-Pbca to 
orthorhombic S-Pbca structure below $\it{T}_{\rm{MI}}$.~\cite{fri2} 
In the latter structure, it is expected that the bandwidth of 
$\it{t}_{\rm{2g}}$ orbitals becomes narrow and the band configuration changes 
to lower the electronic energy.~\cite{nak2,nak3,fri1,ani1} 

In CSRO, bandwidth is mainly controlled by substitution of Sr for Ca. 
The other essential procedure to modify the quasi-two-dimensional 
antiferromagnetic (AF) Mott insulator is filling control, 
by which high temperature superconductivity of cuprates appears.
However, we should take into account that ruthenates possess 
triply degenerate $\it{t}_{\rm{2g}}$ orbitals $4d_{xy}, 4d_{yz}, 4d_{zx}$; 
in contrast, cuprates possess only single orbital 3$d_{x^{2}-y^{2}}$. 
Therefore, it is highly interesting to investigate how this multi-band 
structure of ruthenates would give rise to a different 
characteristic in M/NM transition by filling control. 

It is natural to try substituting trivalent La for divalent Ca 
in order to alter the number of 4$d$ electrons, 
since the ionic radius of La$^{3+}$ ions ($r_{\rm{La}}=1.216\; \rm{\AA}$) 
is between those of Ca$^{2+}$ ions and Sr$^{2+}$ ions. 
Accordingly, we grew single crystals of electron-doping system 
Ca$_{2-\it{x}}$La$_{\it{x}}$RuO$_{4}$ (CLRO)
by a floating zone (FZ) method for the first time. 
It is vital to investigate these quasi-two-dimensional materials by using 
single crystals because of their strongly anisotropic physical properties. 

In this paper we present the phase diagram of CLRO 
obtained from in-plain resistivity, specific heat and dc magnetization. 
CAF and non-metallic (NM) region exists 
just near the parent material Ca$_{2}$RuO$_{4}$ ($0.00 < x < 0.15$). 
The physical characteristics of this region is quite similar to 
those reported in CSRO ($0.00 < x < 0.20$).~\cite{nak2,nak3} 
Metallic ground state appears with increasing $x$ ($x \geq 0.15$). 
Nevertheless, the characteristics of the metallic region just above $x$ = 0.15 
is different from those of CSRO ($0.20 \leq x < 0.50$). 
It is rather similar to those of CSRO with $x \geq 0.50$. 
We discuss the origin of such electronic correlation in the metallic region. 

Recently Cao $et\; al.\;$ also studied this CLRO system, but by using 
single crystals prepared by a flux method using Cl-flux.~\cite{cao2} 
Physical properties of crystals prepared by a flux method often qualitatively 
differ from the intrinsic ones obtained by single crystals grown 
by a FZ method or by polycrystals prepared 
by a conventional solid-state reaction.~\cite{nak1,cao1,bra1,ale1,fuk1,fzf1} 
Comparing with the results reported by Cao $et\; al.\;$, 
we clarify the intrinsic properties of CLRO.

\section{Experimental}
We have recently succeeded in growing single crystals of 
Ca$_{2-\it{x}}$La$_{\it{x}}$RuO$_{4}$ 
($0.00 \leq x \leq 0.20$) 
by a FZ technique with an infrared image furnace.~\cite{nic1}
Typical condition of the single crystal growth has already been 
described.~\cite{fuk1} 

We estimated the concentration of La in as-grown crystals of CLRO 
by energy dispersive x-ray (EDX) analysis : $x$ = 0.00(1), 0.02(1), 0.06(2), 
0.11(2), 0.18(2) and 0.18(2) corresponding to the nominal concentration of 
$x$ = 0.00, 0.015, 0.05, 0.10, 0.15 and 0.20 
in the polycrystalline feed rods used for single crystal growth. 
Although the results by EDX analysis for the nominal concentrations of 
$x$ = 0.15 and 0.20 coincide with each other, 
we designate $x$ as the concentration in the polycrystalline feed rods 
since the other physical properties of the single crystals did 
systematically change with the increase of nominal concentration. 

Braden $et\; al.\;$ reported that after annealing treatment in oxygen, 
polycrystalline Ca$_{2}$RuO$_{4+\delta}$ with $\delta = 0.07(1)$ 
exhibits MI transition at about 150 K.~\cite{bra1} 
In order to investigate the intrinsic properties of CLRO, 
we determined the amount of oxygen $\delta$ in 
Ca$_{2-x}$La$_{x}$RuO$_{4+\delta}$ 
by thermogravimetric analysis (TGA). 
We found that $\delta$ for $x$ = 0.00 and 0.20 were 
0.00(1) and 0.01(1), respectively. 
These results ensure that all the as-grown crystals of CLRO 
are stoichiometric within experimental accuracy. 
Therefore, we may safely assume that the number of 4$d$ electrons 
in the $t_{\rm{2g}}$ orbitals increases with $x$. 
In this paper, we will refer to stoichiometric Ca$_{2}$RuO$_{4}$ 
as S-CRO and oxidized Ca$_{2}$RuO$_{4}$ (Ca$_{2}$RuO$_{4+\delta}$) as O-CRO. 

\begin{figure}
\leavevmode
\epsfxsize=85mm
\epsfbox{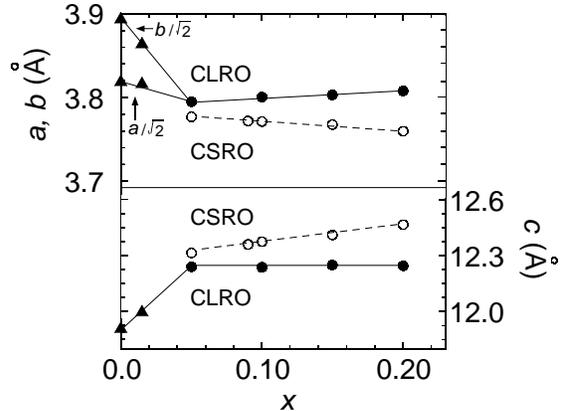}
\caption{
Lattice parameters of Ca$_{2-\it{x}}$La$_{\it{x}}$RuO$_{4}$ 
(solid triangles: $x$ = 0.00 and 0.015, 
solid circles: 0.05 $\leq$ $x$ $\leq$ 0.20) 
at room temperature. 
We have divided the in-plane parameters of the orthorhombic samples 
by $\sqrt{2}$. 
The lattice parameters of Ca$_{2-\it{x}}$Sr$_{\it{x}}$RuO$_{4}$ 
(open circles) are also shown for comparison.~\cite{nak3} 
The change from orthorhombic structure (triangles) 
to tetragonal structure (circles) coincides with that of the 
transport properties from non-metallic behavior to metallic one with $x$.
}
\label{fig:1}
\end{figure}

We examined the structure of the as-grown samples at room temperature 
by x-ray diffraction of powdered single crystals. 
For $x$ = 0.00 and 0.015, the x-ray patterns agreed with 
that of orthorhombic structure. 
For the other $x$, the x-ray pattern fit well to tetragonal structure. 
In Fig.~\ref{fig:1} we summarize 
the lattice parameters of CLRO and CSRO.~\cite{nak3} 
Here, we divide $a$ and $b$ of the orthorhombic samples 
by $\sqrt{2}$ for the sake of convenience. 
As we will describe below, orthorhombic samples exhibit non-metallic behavior; 
however, tetragonal samples exhibit metallic behavior. 
In tetragonal phase, $a$ (= $b$) is longer than that for CSRO 
with corresponding $x$, while $c$ is shorter. 
With increasing $x$, $a$ increases slightly while 
$c$ remains nearly unchanged. 

We measured the in-plane resistivity by a standard four-probe method below 
355 K and the specific heat by a relaxation method between 1.8 and 300 K. 
We investigated the dc magnetization with a commercial SQUID magnetometer 
between 1.8 and 700 K.~\cite{qua1}

\section{Results}

\subsection{Phase diagram}
\begin{figure}
\leavevmode
\epsfxsize=85mm
\epsfbox{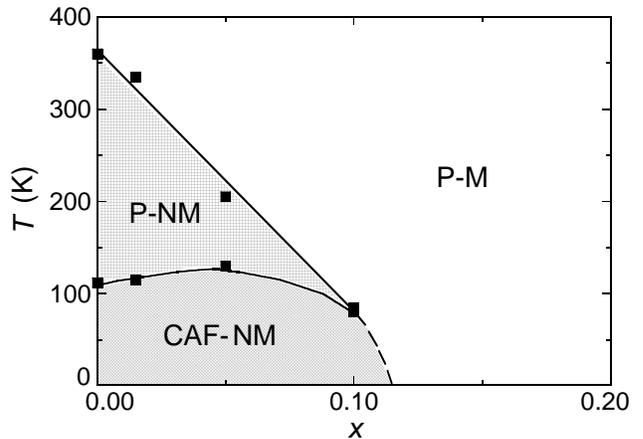}
\caption{
Phase diagram of Ca$_{2-\it{x}}$La$_{\it{x}}$RuO$_{4}$. 
P: paramagnetic, CAF: canted antiferromagnetic, -M: metal, and -NM: non-metal. 
The M/NM transition is associated with the first-order structural transition. 
}
\label{fig:2}
\end{figure}

First, we summarize the physical properties of 
CLRO (0.00 $\leq$ $x$ $\leq$ 0.20) 
in the phase diagram in Fig.~\ref{fig:2}. 
In this phase diagram, we denote the transition temperatures 
observed on heating. 
Instead of M/NM transition temperature for $x$ = 0.015, we plot 
the first-order phase transition temperature of the magnetization, since 
we did not extend the resistivity measurement to high enough temperature. 
Indeed the first-order phase transition temperature of the magnetization 
coincides with MI or M/NM transition temperature 
for $x$ = 0.00, 0.05 and 0.10. 

In the non-metallic region, CAF ordering occurs 
below $\it{T}_{\rm{N}}$. 
This temperature is clearly less than 
the M/NM transition temperature $\it{T}_{\rm{M/NM}}$. 
$\it{T}_{\rm{M/NM}}$ approaches $\it{T}_{\rm{N}}$ with increasing $x$ and 
both vanish below $x$ = 0.15. 
We draw the expected phase boundary with a broken line in Fig.~\ref{fig:2}. 
In the metallic region, apparent Curie-Weiss law is realized. 

In this chapter, we describe the in-plane resistivity, specific heat and 
dc magnetization, which serve the basis for the phase diagram.

\subsection{In-plane resistivity}
\begin{figure}
\leavevmode
\epsfxsize=85mm
\epsfysize=70mm
\epsfbox{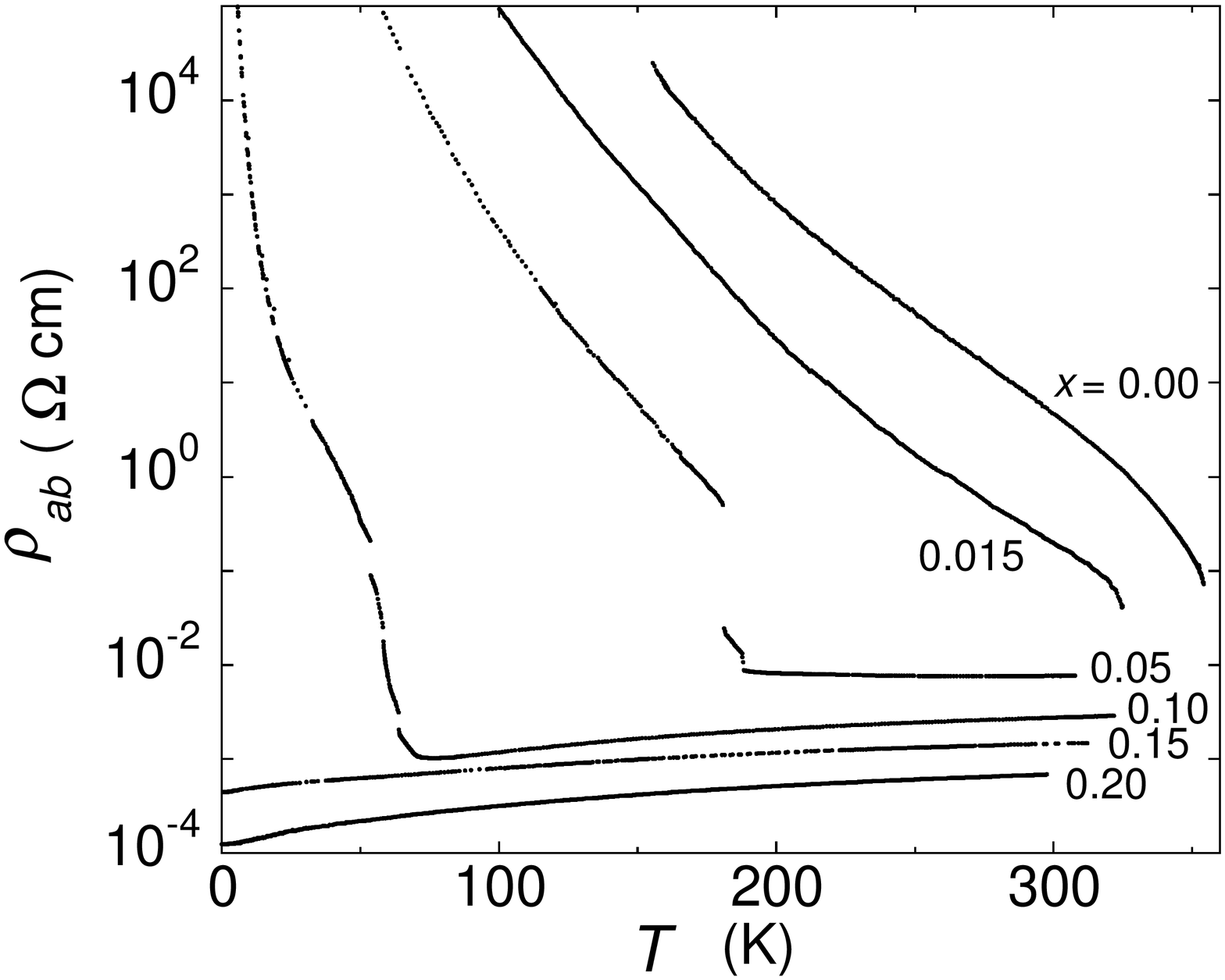}
\caption{The in-plane resistivity $\rho _{ab}$($T$) 
of CLRO (0.00 $\leq$ $x$ $\leq$ 0.20) below 355 K. 
The metal/non-metal transition temperature decreases with $x$. 
Metallic ground state appears at $x$ = 0.15.
}
\label{fig:3}
\end{figure}

In Fig.~\ref{fig:3}, we show the in-plane resistivity 
$\rho _{ab}$($T$) of CLRO below 355 K. 
The data plotted were obtained on cooling below room temperature and 
on heating above room temperature in helium atmosphere. 
We measured the dc resistivity down to 4.2 K for all the samples and the 
ac resistivity down to 0.5 K for $x$ = 0.15 and down to 50 mK for $x$ = 0.20. 

Above 355 K, the measurement had to be terminated for S-CRO 
since the samples shattered at this temperature, 
owing to the first-order structural phase transition. 
This transition nearly coincides with the insulator-"near" metal transition 
temperature of the Ca$_{2}$RuO$_{4}$ crystals 
prepared by a flux-method.~\cite{ale1} 
The resistivity exhibits insulating behavior on cooling 
down to about 150 K. 
Below 150 K, the $\rho _{ab}$ became too large 
for our present measurement setup. 
Nevertheless we confirmed that below this temperature the resistivity 
remains higher than the value at 150 K. 

For $x$ = 0.015, the crystals did not shatter at least up to 325 K. 
For $x$ = 0.05 and 0.10, the $\rho _{ab}$($T$) exhibits 
metallic behavior on cooling but suddenly increases 
by two orders of magnitude at 188 K ($x$ = 0.05) and at 77 K ($x$ = 0.10). 
Below each characteristic temperature, 
the $\rho _{ab}$ becomes non-metallic and 
monotonously increases beyond $10^{5}$ $\rm{\Omega cm}$ on cooling. 
On warming, the crystals again exhibit the hysteretic M/NM transition 
at 206 K ($x$ = 0.05) and at 84 K ($x$ = 0.10). 
Hence, the M/NM transition in CLRO is most probably of the first order. 
The resistivity on warming above $\it{T}_{\rm{M/NM}}$ does not recover 
the same value that was obtained on cooling. 
This is clearly due to micro cracks, 
actually found after the measurement, which are 
most likely introduced by the first-order structural phase transition. 
From this result we deduce that the M/NM transition coincides with 
the first-order structural phase transition in CLRO, as reported 
in the parent material S-CRO and CSRO.~\cite{ale1,nak2,fri1} 
Cao $et\; al.\;$ also reported the same phenomenon.~\cite{cao2} 
However, we notice that the resistivity at low temperatures show no saturation 
behavior, in contrast with the report by Cao $et\; al.\;$~\cite{cao2} 

For $x$ = 0.15 and 0.20, the $\rho _{\rm{ab}}$($T$) almost linearly decreases 
on cooling down to 1.0 K ($x$ = 0.15) and 0.5 K ($x$ = 0.20). 
The resistivity saturates below 1.0 K ($x$ = 0.15) and 0.5 K ($x$ = 0.20) 
with the residual resistivities of $\rho_ {0}$ = 
$4.5\times10^{2}\;\rm{\mu\Omega cm}$ ($x$ = 0.15) 
and $\rho_ {0}$ = $1.3\times10^{2}\;\rm{\mu\Omega cm}$ ($x$ = 0.20). 
We do not find any sign of superconductivity 
at least down to 0.5 K ($x$ = 0.15) and 50 mK ($x$ = 0.20) so far. 

\subsection{Specific heat}
\begin{figure}
\leavevmode
\epsfxsize=85mm
\epsfbox{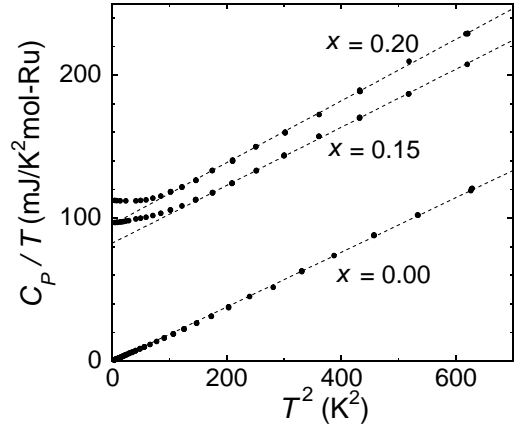}
\caption{Electronic specific heat $\it{C_{P}/T}$ vs $\it{T}^{\rm{2}}$ of 
Ca$_{2-\it{x}}$La$_{\it{x}}$RuO$_{4}$.
}
\label{fig:4}
\end{figure}

In Fig.~\ref{fig:4}, we plot the specific heat divided by temperature, 
$\it{C_{P}/T}$, for $x$ = 0.00, 0.15 and 0.20 
against $\it{T}^{\rm{2}}$ below 25 K. 
We could not measure the specific heat of the crystals for $x$ = 0.05 and 0.10 
since the crystals attached to the sapphire plate with vacuum grease 
always came off on cooling through $\it{T}_{\rm{M/NM}}$. 

We performed fitting the data to the following formula 
below 25 K for $x$ = 0.00 and between 10 and 25 K for $x$ = 0.15 and 0.20: 
\begin{eqnarray}
\it{C_{P} / T} = \gamma + \beta \it{T}^{2}. 
\label{fml:2}
\end{eqnarray}
Electronic specific heat coefficient $\gamma$ directly reflects 
the density of states (DOS) at the Fermi level, 
$\it{D}\rm{(}\varepsilon_{\rm{F}}\rm{)}$. 
From $\beta$, we can deduce the Debye temperature by 
${\it{\Theta}}_{\rm{D}} = (\rm{12\pi}^{4}\it{N}k_{\rm{B}}/\rm{5}\beta)^{1/3}$
, where $\it{N}$(= 7$\it{N}_{\rm{A}}$: $\it{N}_{\rm{A}}$ is the Avogadro 
number) is the number of atoms per mole and $k_{\rm{B}}$ is 
the Boltzmann constant. 
We summarize the parameters ($\gamma$, ${\it{\Theta}}_{\rm{D}}$)
obtained from the data in Table~\ref{table:1}. 

The result that $\gamma$ is equal to $0.0 \pm 0.2\; \rm{mJ/ K^{2}}$mol-Ru 
for S-CRO indicates an energy gap at the Fermi level in the ground state. 
Hence, S-CRO is an insulator. 

$\gamma$ for the metallic samples is by a factor of more than two larger than 
that of the spin-triplet superconductor Sr$_{2}$RuO$_{4}$, 
37.5 mJ/K$^{2}$mol-Ru.~\cite{mae2,nis1} 
Slight deviation from eq.~(\ref{fml:2}) below 10 K 
might reflect strong magnetic fluctuation. 
All the values of ${\it{\Theta}}_{\rm{D}}$, including that of 
Sr$_{2}$RuO$_{4}$ (${\it{\Theta}}_{\rm{D}} = 410 \pm 10 \rm{K}$), 
are almost the same.~\cite{mac2} 

\begin{table}
\caption{
Electronic specific heat coefficient $\gamma$ and 
Debye temperature $\it{\Theta _{D}}$ of Ca$_{2-\it{x}}$La$_{\it{x}}$RuO$_{4}$.
}
\label{table:1}
\begin{tabular}{@{\hspace{\tabcolsep}\extracolsep{\fill}}ccc} \hline
$x$ & $\gamma$ (mJ/K$^{2}$mol-Ru) & $\it{\Theta _{D}}$ (K) \\ \hline
0.00 & 0.0(2) & 400 \\
0.15 & 82(1) & 420 \\
0.20 & 95(1) & 420 \\ \hline
\end{tabular}
\end{table}

\subsection{Magnetization} 

\subsubsection{Magnetization of S-CRO}
\begin{figure}
\leavevmode
\epsfxsize=85mm
\epsfbox{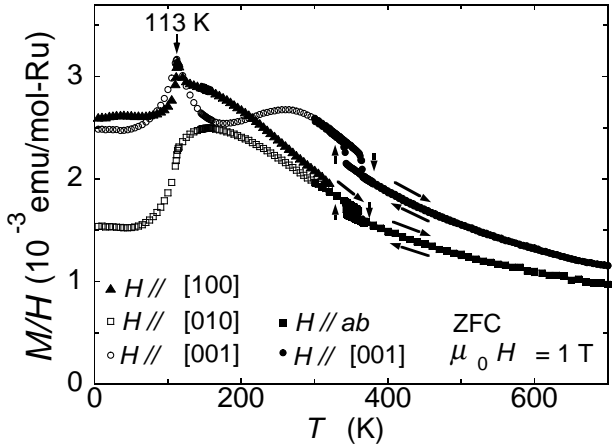}
\caption{
The magnetic susceptibilities $\chi$($\it{T}$) of stoichiometric 
Ca$_{2}$RuO$_{4}$ (S-CRO). 
S-CRO exhibits CAF ordering below 113 K. 
The hysteretic behavior around 350 K corresponds to 
metal-insulator transition, which coincides 
with the first-order structural phase transition.~\cite{ale1} 
}
\label{fig:5}
\end{figure}

In Fig.~\ref{fig:5}, the anisotropic dc magnetic susceptibilities, 
$\it{M}(\it{T})/\it{H} \equiv \chi (\it{T})$ 
($\it{\mu}_{\rm{0}}\it{H}$ = 1 T), in zero-field-cooled (ZFC) 
sequence are shown for S-CRO. 
Below 320 K, we used a uniaxial sample rotator. 
The directions of the applied field were along the [100], [010] and [001] 
axes of the orthorhombic structure with Pbca symmetry. 
By adjusting the [001] axis to be parallel to the rotational axis, 
we applied the magnetic field parallel to the [100] axis and the [010] axis. 
Between 320 and 700 K, we measured the $\chi (\it{T})$ along the [001] axis 
and along one of the axes in the $ab$ plane on heating and on cooling. 

We found slight hysteresis between the data of ZFC sequence and 
field-cooled (FC) sequence below about 160 K. 
However, the magnitude of this hysteresis is 
negligible compared with that of O-CRO as we show in Fig.~\ref{fig:6}. 
Therefore, let us concentrate on the ZFC data of S-CRO. 

The prominent features in Fig.~\ref{fig:5} are sharp cusps of $\chi$ 
along the [100] and [001] axes at 113 K. 
Below this temperature, $\chi$ along each of the three axes 
decreases rapidly on cooling. 
This is ascribable to CAF ordering 
below $\it{T}_{\rm{N}} = \rm{113\;K}$.~\cite{fuk1} 

Between $\it{T}_{\rm{N}}$ and 320 K, $\chi$ along each axis cannot 
be fitted to the Curie-Weiss law: 
\begin{eqnarray}
\chi (\it{T}) & = & 
\chi_{\rm{p}} + \frac{\it{C}}{\it{T}-\it{\Theta}_{\rm{CW}}}, \label{fml:3}
\end{eqnarray}
Here, $\chi_{\rm{p}}$ represents the paramagnetic component 
($\it{eg.}$ Pauli paramagnetism and van Vleck orbital paramagnetism, 
as well as diamagnetic susceptibility of the ion cores). 
$\it{C}$ is the Curie constant and $\it{\Theta}_{\rm{CW}}$ is 
the Curie-Weiss temperature. 
It is surprising that $\chi_{[001]}$($\it{T}$) has a broad peak 
at around 260 K since no static magnetic ordering was found 
by neutron diffraction.~\cite{bra1} 
This temperature is substantially higher than $\it{T}_{\rm{N}}$; 
we cannot attribute this to the influence of CAF correlation. 

On heating $\chi_{ab}$ and $\chi_{c}$ abruptly decrease at 360 K 
corresponding to $\it{T}_{\rm{MI}}$. 
On cooling $\chi_{ab}$ and $\chi_{c}$ recover the value 
in the insulating region at 344 K. 
This hysteretic behavior of the magnetization indicates 
the phase transition of the first order. 
$\chi_{ab}$ and $\chi_{c}$ apparently follow the Curie-Weiss law 
between 360 and 700 K, 
$\it{\Theta}_{\rm{CW}} = \rm{-78\; K}$ parallel to the $ab$ plane 
and $\rm{-73\; K}$ parallel to the $c$ axis. 
The total spin $\it{S}$ derived by $\it{C} = {\it{N}}_{\rm{A}}g^{\rm{2}}\it{S}\rm{(}\it{S}\rm{+1)}{{\mu}_{\rm{B}}}^{\rm{2}} / \rm{3}k_{\rm{B}}$ 
(${\mu}_{\rm{B}}$: the Bohr magneton) 
is equal to 0.8 parallel to the $ab$ plane and 0.9 parallel to the $c$ axis, 
reflecting the nearly localized moment $\it{S} \simeq \rm{1}$. 
The other parameter $\chi_{\rm{p}}$ is consistent with that obtained 
from polycrystalline samples and its magnitude is negligible.~\cite{nak3} 

\subsubsection{Magnetization in the non-metallic region}
\begin{figure}
\leavevmode
\epsfxsize=85mm
\epsfbox{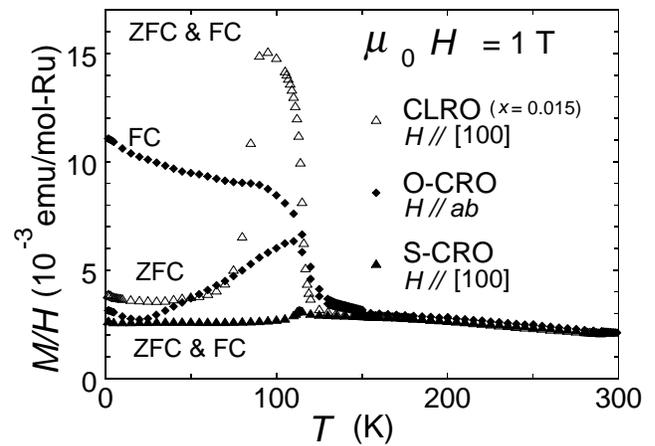}
\caption{
The $\chi_{\rm{[100]}}$($\it{T}$) of S-CRO and 
CLRO ($x = 0.015$) and the $\chi_{ab}$($\it{T}$) of 
oxidized Ca$_{2}$RuO$_{4}$. 
Slight carrier-doping induces ferromagnetic component 
in the magnetically ordered region. 
}
\label{fig:6}
\end{figure}

In Fig.~\ref{fig:6}, we plot the magnetic susceptibility 
along the [100] axis for S-CRO and CLRO ($x$ = 0.015). 
We add the in-plane magnetic susceptibility of single crystals of O-CRO, 
which was prepared by annealing the single crystals of S-CRO 
under 12.0 MPa of oxygen at 540 $^{\circ}$C for 90 hours. 
The single crystals of O-CRO exhibited M/NM transition at 290 K on cooling 
but shattered at 320 K on heating (not shown). 
We could not determine the amount of oxygen $\delta$ 
in Ca$_{2}$RuO$_{4+\delta}$ by TGA, since the mass of single-phase crystals 
was not large enough for high-precision TGA. 
Thus, we estimated $\delta \sim 0.04$ 
from the difference of the mass before and after annealing. 
This estimation is consistent with the result that $\it{T}_{\rm{M/NM}}$ of 
single-crystalline O-CRO, $\sim$ 320 K, is between that of S-CRO 
with $\delta$ = 0.00(1), $\sim$ 350 K, and that of polycrystalline O-CRO 
with $\delta$ = 0.07(1), $\sim$ 150 K.~\cite{bra1} 
We may regard O-CRO as another filling control system derived from S-CRO: 
a hole-doping system. 

From the large magnetization and its hysteresis shown in Fig.~\ref{fig:6}, 
we may conclude that ferromagnetic (FM) component appears in the AF ground 
state of S-CRO once carriers are introduced into the $t_{\rm{2g}}$ orbitals. 
Despite the prominent peak along the [100] axis of CLRO ($x$ = 0.015), 
$\chi$($\it{T}$) along either the [010] axis or the [001] axis 
does not show any corresponding peak (not shown). 
Instead the $\chi_{[010]}$ and the $\chi_{[001]}$ of CLRO ($x$ = 0.015) 
are similar to those of S-CRO in magnitude. 
Static AF ordering is observed by neutron diffraction in S-CRO and 
in polycrystalline O-CRO with $\delta = 0.07(1)$ below 150 K 
in spite of the presence of finite FM component.~\cite{bra1} 
This result supports that the magnetic ordering of single-crystalline O-CRO 
below 160 K is also ascribable to AF ordering with FM component. 

For CLRO ($x$ = 0.015), we found the hysteretic change of $\chi$ 
at 334 K on heating and at 323 K on cooling (not shown). 
This behavior is essentially the same as that seen in S-CRO at around 350 K. 
The crystals after the measurement had shattered. 
We infer that this first-order phase transition comes from the 
structural phase transition, which coincides with M/NM transition. 

\begin{figure}
\leavevmode
\epsfxsize=85mm
\epsfbox{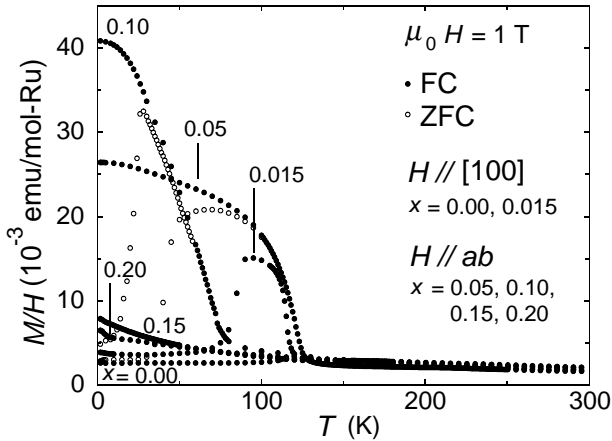}
\caption{
The $\chi_{\rm{[100]}}$($\it{T}$) for $x$ = 0.00 and 0.015 and the 
$\chi_{ab}$($\it{T}$) for $0.05 \leq x \leq 0.20$. 
In the non-metallic region, canted antiferromagnetic ordering 
occurs below $\it{T}_{\rm{N}}$. 
Apparent Curie-Weiss behavior was found for $x$ = 0.15 and 0.20 down to 1.8 K. 
}
\label{fig:7}
\end{figure}

In Fig.~\ref{fig:7}, we plot the magnetic susceptibilities 
$\chi_{\rm{[100]}}$($\it{T}$) for $x$ = 0.00 and 0.015 and 
the $\chi_{ab}$($\it{T}$) for $0.05 \leq x \leq 0.20$ in ZFC and FC sequences. 
Only the data for $x$ = 0.05 and 0.10 show the strongly hysteretic behavior 
below 170 K ($x$ = 0.05) and 75 K ($x$ = 0.10). 

As also reported by Cao $et\; al.\;$, the FM component emerges 
gradually with $x$.~\cite{cao2} 
We attribute this to the alteration of the interlayer magnetic mode 
of CAF ordering as we will discuss in \S 4. 2. 3. 

\subsubsection{Magnetization in the metallic region}
In the metallic region, there is no sign of magnetic ordering. 
Instead, apparent Curie-Weiss law is realized down to low temperatures. 
In this subsection, we describe the magnetization of CLRO 
for $x$ = 0.15 and 0.20. 

We fit the data to eq.~(\ref{fml:3}) above 100 K. 
We obtain the negative Curie-Weiss temperature and 
finite paramagnetic component $\chi_{\rm{p}}$
(= 5-6 $\times$10$^{-4}$ emu/mol-Ru). 
Since we consider the conducting electrons all originating from 
the same $t_{\rm{2g}}$ orbitals, it is not natural to expect 
the two magnetic species: the nearly localized moment 
giving the Curie-Weiss term and the itinerant electrons 
giving the paramagnetic component $\chi_{\rm{p}}$ in eq.~(\ref{fml:3}). 
(An alternative interpretation is recently given 
by Anisimov $et\; al.\;$~\cite{ani1})

In the presence of strong correlation and spin fluctuation 
among itinerant electrons, it is expected that the Pauli susceptibility 
is modified to $\chi_{\rm{P}}$($\it{T}$) given by ~\cite{mor1} 
\begin{eqnarray}
\chi_{\rm{P}}(\it{T}) & = & \frac{\chi_{\rm{0}}(\it{T})}
{1 - \alpha \chi_{\rm{0}}(\it{T}) + \lambda (\it{T})}\, , \\
\chi_{\rm{0}}(\it{T}) & = & \chi_{\rm{0}}(1 + c_{0}{\it{T}}^{2}) . \nonumber 
\label{fml:5}
\end{eqnarray}
Here, $\alpha$ is the correlation energy, 
$\chi_{0}$ is the Pauli susceptibility of non-interacting system, 
$c_{0}$ is the coefficient depending on DOS at around the Fermi level 
and is ordinarily proportional to $\it{T}_{\rm{F}}^{\rm{-2}}$, 
$\lambda$($\it{T}$) is proportional to $T$ in the presence of FM fluctuation. 
Although ${\chi_{\rm{P}}}^{\rm{-1}}(\it{T})$ is approximately 
proportional to $a_{\rm{0}}\it{T}+b_{\rm{0}}$ ($a_{\rm{0}},\, b_{\rm{0}} > 0$) 
which gives apparent negative Curie-Weiss temperature, 
it exhibits the slight curvature of 
$\rm{d}^{2}({\chi_{\rm{P}}}^{\rm{-1}}) \big/ \rm{d}\it{T}^{\rm{2}} > \rm{0}$ 
or $< \rm{0}$, 
since $\chi_{\rm{0}}(\it{T})$ contains the $c_{0}\it{T}^{\rm{2}}$ term. 
In fact, ${\chi_{ab}}^{\rm{-1}}$($\it{T}$) for $x$ = 0.15 and 0.20 gives 
the negative Curie-Weiss temperature and 
exhibits such temperature dependence above 1.8 K. 
Consequently, the experimental ${\chi_{ab}}$($\it{T}$) 
for $x$ = 0.15 and 0.20 is well ascribed as 
the Pauli susceptibility in the presence of FM fluctuation. 

Although it is not a simple Fermi liquid, 
it is worthwhile to evaluate the Wilson ratio $\it{R}_{\rm{W}}$ 
in order to assess the magnetic correlation among the itinerant electrons. 
$\it{R}_{\rm{W}}$ is given by the following formula: 
\begin{eqnarray}
\it{R}_{\rm{W}} & = & \frac{\chi_{\rm{P}}\rm{(0\;K)}\;/\;\chi_{0}}
{\gamma\;/\;\gamma_{0}}\nonumber \\
& = & \frac{\rm{4}{\rm{\pi}}^{2} {k_{\rm{B}}}^{2} \chi_{\rm{P}}\rm{(0\;K)}}
{\rm{3}{(g\mu_{\rm{B}})}^{2}\gamma} \; . 
\label{fml:4}
\end{eqnarray}
Here, $\chi_{\rm{P}}\rm{(0\;K)}$ is 
the experimental Pauli susceptibility in the limit of 0 K, 
$\gamma$ is the electronic specific heat coefficient obtained from experiment, 
and $\gamma_{0}$ is that of non-interacting system. 

To evaluate $\it{R}_{\rm{W}}$, we use the experimental value 
${\chi_{ab}}$(2 K) to approximate $\chi$(0 K), 
the electronic specific heat coefficient $\gamma$ in table~\ref{table:1}. 
Wilson ratio $\it{R}_{\rm{W}}$ are 6.9 for $x$ = 0.15 and 4.8 for $x$ = 0.20. 
$\it{R}_{\rm{W}}$ for both samples is considerably larger than the value one 
or two, which are expected for a non-interacting electron system or for a 
strongly correlated electron system, respectively. 
This result is in accord with the existence of 
additional FM fluctuation among the itinerant electrons 
over the mass enhancement at least for $x$ = 0.15 and 0.20. 
These magnetic characteristics in the metallic region of CLRO are similar 
to those reported in CSRO for $x \geq 0.50$.~\cite{nak2,nak3}

\section{Discussion}

\subsection{Mott insulator Ca$_{2}$RuO$_{4}$}
In our previous papers we have revealed that S-CRO is a Mott insulator 
on the basis of the results with single crystals grown by a FZ method 
and polycrystalline samples made by a conventional solid state 
reaction.~\cite{nak1, bra1, fuk1} 
However, Cao $et\; al.$ had implied that Ca$_{2}$RuO$_{4}$ is 
a localized metal from the results of the flux-grown crystals, 
since $\gamma$ is about $4\; \rm{mJ/ K^{2}}$mol-Ru and resistivity exhibits 
a variable-range-hopping (VRH) behavior.~\cite{cao1} 
In addition Mazin and Singh also agreed with this conclusion 
for the reason that a small Fermi surface exists in S-CRO 
in their local density approximation calculation.~\cite{maz1} 
Although $\gamma$ is still finite, Cao and Alexander $et\; al.\;$ 
have recently agreed with our previous claim and stated that Ca$_{2}$RuO$_{4}$ 
is not a localized metal but a Mott insulator.~\cite{ale1,cao2} 
Mazin and Singh have also agreed with the fact,~\cite{maz2} since the 
intrinsic $\gamma$ of S-CRO is $0.0(2)\; \rm{mJ/ K^{2}}$mol-Ru, which is 
obtained from high quality single crystals grown by a FZ method.~\cite{fuk1} 
We note that experimental finite $\gamma$ is attributable to the off 
stoichiometry of the single crystals, since Ca$_{2}$RuO$_{4}$ is 
quite sensitive to annealing treatment. 

S-CRO exhibits CAF ordering below 113 K. 
The associated spin canting is ascribable to the Dzyaloshinsky-Moriya 
(DM) interaction.~\cite{nak1,fuk1} 
This suggests that spin-orbit coupling is important in S-CRO. 
According to the spin-resolved circularly-polarized photoemission 
and O $1s$ x-ray absorption spectroscopy (XAS) on our single crystals of S-CRO 
performed by Mizokawa $et\; al.\;$, the spin-orbit coupling 
indeed plays an important role in S-CRO.~\cite{miz1} 
Here, let us discuss the origin of the broad peak of $\chi_{[001]}$($\it{T}$) 
of S-CRO at around 260 K (Fig.~\ref{fig:4}). 
Mizokawa $et\; al.$ showed that the orbital angular moment 
exists along the [001] axis at 300 K, but on cooling to 90 K, 
it alters its direction to within the $ab$ plane. 
In addition, they showed, by calculation, that the form of the RuO$_{6}$ 
octahedra in S-CRO dictates the direction of the orbital moment; 
orbital moment is along the [001] axis for elongated and 
regular RuO$_{6}$ octahedra, while perpendicular to the [001] axis 
for flattened RuO$_{6}$ octahedron. 
Neutron diffraction on polycrystalline S-CRO has revealed that the RuO$_{6}$ 
octahedra is elongated above 300 K, regular at around 300 K 
and flattened below 300 K.~\cite{bra1,fri1} 
We expect the magnetic anisotropy to be influenced by changes 
in the direction of the orbital moment, as well as 
by the associated polarization of the spin moment. 
The experimental result that $\overline{\chi_{ab}}$
$\big( \equiv (\chi_{[100]}+\chi_{[010]})/2\, \big)$ is smaller 
than $\chi_{[001]}$ above 200 K but becomes larger than $\chi_{[001]}$ between 
150 and 200 K (Fig.~\ref{fig:5}) supports this interpretation.~\cite{fuk2} 

\subsection{Non-metallic region}

\subsubsection{Origin of non-metallic ground state}
In the parent material S-CRO, the metal to insulator transition coincides 
with the first-order structural phase transition of 
the quasi-tetragonal L-Pbca phase to the orthorhombic S-Pbca phase 
as we noted in \S 1.~\cite{ale1,fri1} 
In CSRO and the hole-doping system O-CRO, it is revealed 
by neutron diffraction on polycrystalline samples 
that the situation is almost the same.~\cite{bra1,nak3,fri1} 
In CLRO investigated here, non-metallic crystals 
have the orthorhombic structure whereas metallic crystals have 
the tetragonal structure at room temperature. 
Furthermore, M/NM transition in this system is also of the first order. 
These results indicate that the metal to non-metal transition coincides 
with the first-order structural phase transition between the high-temperature 
L-Pbca phase to the low-temperature phase in CLRO. 
Neutron diffraction experiment is necessary to confirm that 
the low-temperature structure is indeed S-Pbca even for $x \neq 0.00$. 

The change from the L-Pbca phase to the S-Pbca phase is 
characterized by the flattening of the RuO$_{6}$ octahedra. 
The lattice parameters $a$ and $b$ are elongated and $c$ is shrunk 
reflecting the flattening. 
The flattening of the RuO$_{6}$ octahedra is expected to greatly influence 
the crystal field which mainly determines the electron occupancy 
in the triply degenerate orbital state of ruthenate. 
By the flattening, nearly degenerate $t_{\rm{2g}}$ orbitals 
is expected to form the lowest $4d_{xy}$ orbital and the lower and 
the upper Hubbard band which originates from the nearly doubly degenerate 
$4d_{yz}$ and $4d_{zx}$ orbitals. 
O $1s$ XAS on S-CRO has clarified that the $4d_{xy}$ orbital is energetically 
lower than the other $4d_{yz}$ and $4d_{zx}$ orbitals at 90 K, 
supporting our expectation.~\cite{miz1} 
Such structural change is attributable to the Jahn-Teller effect.~\cite{yos1} 

The physical properties in the non-metallic region of CLRO is 
qualitatively quite similar to those of the Mott insulator S-CRO. 
Hence, we infer that the ground state of CLRO in the non-metallic region 
is in a broad sense an Mott insulator. 
However, this non-metallic phase is more appropriately described by the 
Anderson localization of the excess $4d$ electrons 
introduced by La substitution. 

We did not find the saturation behavior of the resistivity at low temperatures 
in the non-metallic region for any $x$, in contrast with the observation 
in the flux-grown crystals 
($0.07 \leq x \leq 0.11$: $\it{T}_{\rm{M/NM}} \leq \rm{210\; K}$).~\cite{cao2} 
Such behavior suggests that delocalized electrons exist 
in the flux-grown crystals. 
Nevertheless, magnetic properties of single crystals grown both by a FZ 
method and by a flux method are almost the same. 

\subsubsection{Conduction mechanism}
For ordinary semiconducting materials with band gaps, 
the following formula is often used:~\cite{mot1} 
\begin{eqnarray}
\rho _{ab} \rm{(}\it{T}\rm{)} = 
\it{A}_{\rm{0}}\exp\;\biggl( \frac{\it{\Delta}_{n}}{\it{T}} \biggr) 
^{\frac{\rm{1}}{n\rm{+1}}}  (n = \rm{0, 1, 2, 3}). 
\label{fml:1}
\end{eqnarray}
For $n$ = 0, the formula represents the activation-type conductivity: 
$\it{\Delta}_{\rm{0}}$ is the activation energy. 
For the other $n$, eq.~(\ref{fml:1}) represents the $n$ dimensional 
VRH without interaction among the localized electrons: 
$\it{\Delta}_{n}$ is the characteristic temperature. 
If we consider the Coulomb interaction between localized electrons, 
$n$ is unity for three dimensional system.~\cite{efr1} 
In the VRH mechanism, we should notice that DOS at the Fermi level, 
$\it{D}\rm{(}\varepsilon_{\rm{F}}\rm{)}$, is finite. 

Each $\rho_{ab}$-curve for $0.00 \leq x \leq 0.10$ 
has an inflection point below $\it{T}_{\rm{M/NM}}$ (Fig.~\ref{fig:3}). 
We fit the data of $\rho_{ab}$ below the inflection point and found at least 
$n \neq 0$ for all the samples and $n$ = 1 for $x$ = 0.10. 
Fitting with $n$ = 1, 2 or 3 gave equally satisfactory results 
for $x$ = 0.00, 0.015 and 0.05. 
Cao $et\; al.\;$ reported that they found $n$ equal to 1 for the samples of 
CLRO which did not exhibit saturation behavior of the resistivity.~\cite{cao2} 

In CLRO, La substitution would bring a random 
potential, which could localize the itinerant electrons. 
In fact, $n \neq 0$ indicates that 
conduction mechanism is not the simple activation-type but VRH. 
At the same time, we expect strong electronic correlation 
since the parent material S-CRO is the Mott insulator. 
Thus, $n$ = 1 for $x$ = 0.10 suggests that the so-called soft Coulomb gap 
exists at the Fermi level. 
Consequently, we conclude that conduction mechanism in the non-metallic region 
is VRH with a small gap formed by strong electronic correlation. 

Finally we note that the activation energy 
between 150 and 250 K is about 0.2 eV for S-CRO 
which is consistent with the previous results.~\cite{nak1,cao1,ale1,puc1} 
This activation energy diminishes with increasing $x$. 

\subsubsection{Magnetic correlation}
Neutron diffraction on polycrystalline S-CRO and O-CRO samples revealed that 
there are two kinds of magnetic modes for static AF ordering 
in S-CRO and O-CRO: A-centered and B-centered. 
S-CRO has two coexisting modes: A-centered (major phase) and 
B-centered (minor phase). 
This is in contrast with O-CRO, which has only one mode: B-centered. 
A very small magnetic hysteresis in S-CRO indicates that 
it contains B-centered domains as a minor phase. 

\begin{figure}
\leavevmode
\epsfxsize=85mm
\epsfbox{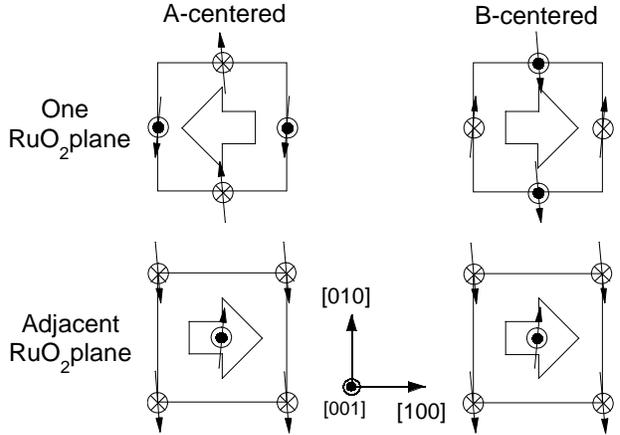}
\caption{
Schematic spin configuration of CAF ordering in S-CRO and O-CRO 
consistent with neutron diffraction and magnetization.~\cite{nak1,bra1,fuk1} 
Solid circles in open circles denote that the [001] component of 
canted moments is positive, while crosses in open circles denote negative. 
In A-centered mode, finite moments induced by spin canting (big arrow) 
cancel out between neighboring RuO$_{2}$ planes. 
In B-centered mode, however, finite moments induce FM component 
along the [100] axis. 
}
\label{fig:8}
\end{figure}

In order to explain AF ordering in S-CRO and O-CRO, 
CAF order has been proposed.~\cite{nak1,bra1,fuk1} 
By taking account of DM interaction associated with spin-orbit coupling, 
we modify the spin arrangements, which are depicted in ref.~\ref{refbra1}, 
consistent with the rotation and tilt configuration of RuO$_{6}$ octahedra 
in Pbca symmetry and with magnetization (Fig.~\ref{fig:8}). 
Nearly the same arrangements have also been discussed 
by Braden $et\; al.$~\cite{bra1} 
However, we propose to include the component of the canted moments 
along the [001] axis, which is necessary to explain the sharp peak 
in $\chi_{[001]}$($\it{T}$) of S-CRO at 113 K (Fig.~\ref{fig:5})~\cite{fuk1} : 
solid circles in open circles denote that the [001] component of 
canted moments is positive, while crosses in open circles denote negative. 
The spin canting will induce finite moment (big arrow)
only along the [100] axis within a RuO$_{2}$ plane. 
These canted moments cancel out between 
neighboring RuO$_{2}$ planes in the A-centered arrangement; 
they add up to finite FM component along the [100] axis in the B-centered one. 
Consequently, three-dimensional AF ordering occurs 
in the A-centered mode, whereas FM component appears 
along the [100] axis in the B-centered AF ordering. 
We attribute magnetic structures in S-CRO and O-CRO to these kinds of 
CAF ordering originating from the two magnetic modes and the spin canting. 

We found a peak in $\chi$($\it{T}$) of CLRO for $x$ = 0.015 only along 
the [100] axis at around 95 K (Fig.~\ref{fig:6}) and no hysteretic behavior 
between ZFC and FC sequences. 
This remarkable temperature dependence is explained if the B-centered mode 
is more stable just below $\it{T}_{\rm{N}}$, though the A-centered mode is 
dominant in the ground state. 
In contrast, $\chi_{ab}$($\it{T}$) of CLRO for $x$ = 0.05 and 0.10 
in both ZFC and FC sequences is qualitatively quite similar to that of O-CRO. 
From these results, magnetically ordered phase of CLRO may also be ascribable 
to CAF ordered one, in which the B-centered mode becomes more stable than 
the A-centered one with increasing $x$ or $\it{T}$. 
This interpretation is in sharp contrast with that by Cao $et\; al.\;$, 
since they claimed that the FM component in the non-metallic region is 
due to the Stoner enhancement originating from the itinerant electrons 
introduced by La substitution.~\cite{cao2} 

\subsection{Metallic region}
The ground state of CLRO for $x$ = 0.15 and 0.20 is metallic. 
Considerably large $\gamma$ suggests that 
finite DOS appears by La substitution and that 
strong electronic correlation exists in the metallic state. 
In this subsection, we discuss 
the physical properties of CLRO for $x \geq 0.15$. 

The resistivity exhibits nearly linear temperature dependence. 
Hence, we may conclude that the ground state is at least not 
a simple Fermi liquid. 
In this region, we did not find any evidence for short-range AF correlation, 
which is the main characteristic of the metallic state adjacent to 
the M/NM boundary in CSRO ($0.20 \leq x < 0.50$).~\cite{nak2,nak3} 
Instead, the magnetic properties of CLRO in this region is 
similar to those in CSRO with $x \geq 0.50$. 
Considerably large Wilson ratio for $x$ = 0.15 and 0.20 
indicates that FM fluctuation is strong in this region. 
The origin of FM correlation may be associated 
with the van-Hove singularity and the resulting Stoner 
enhancement~\cite{ogu1,maz1,nom1}, which itinerant electron system 
in quasi-two-dimensional ruthenates generally exhibit. 

\subsection{Filling control of Ca$_{2}$RuO$_{4}$}
The ground state of CLRO ($0.00 < x < 0.15$) is CAF-NM. 
Almost the same CAF-NM ground state is realized 
in CSRO ($0.00 < x < 0.20$).~\cite{nak2,nak3} 
The important coincidence between the both systems is that 
M/NM transition occurs simultaneously 
with the first-order structural phase transition and that CAF ordering 
occurs at some temperature which is lower than $\it{T}_{\rm{N}}$. 
We concluded that both ground states are in a broad sense 
Mott insulators.~\cite{nak2,nak3} 
These non-metallic ground states are unstable because slight La or Sr 
substitution for Ca will induce metallic ground states. 
Concerning CSRO, only bandwidth of nearly triply degenerate $t_{\rm{2g}}$ 
orbitals is controlled by substituting divalent Sr for equivalent Ca. 
In CLRO, the bandwidth is also influenced 
by substituting trivalent La for heterovalent Ca. 
However, the experimental fact that metallic ground state in CLRO appears 
somewhat above 0.10 less than the corresponding boundary in CSRO 
at $x = 0.20$ indicates that filling control (electron-doping) 
in the $t_{\rm{2g}}$ orbitals is effective 
in destroying the non-metallic ground state, 
since the smaller La$^{3+}$ ions cannot more effectively change 
the bandwidth than the larger Sr$^{2+}$ ions. 
Cao $et\; al.\;$ also noted this from independent results.~\cite{cao2} 
The metallic ground state of CLRO with $x \geq 0.15$ is 
rather different from that of CSRO with $0.20 \leq x < 0.50$ 
but is similar to that of CSRO with $x \geq 0.50$. 
This result also implies that the filling control changes 
the parent compound more rapidly. 

Unlike cuprates, the first-order M/NM transition occurs 
simultaneously with the structural phase transition in CLRO, 
indicating the importance of Jahn-Teller effect. 
This effective filling control through the \it{orbital degeneracy tuning} 
\rm{brings} a metallic phase with FM fluctuation in CLRO. 
This is quite in contrast with the filling control of half-filled 
single orbital, which brings a metallic (or superconducting) phase 
with AF fluctuation in cuprates.

\section{Conclusion}
We have succeeded in growing single crystals of electron doping system 
Ca$_{2-\it{x}}$La$_{\it{x}}$RuO$_{4}$ (0.00 $\leq$ $x$ $\leq$ 0.20) 
by a FZ method. 
We have proposed the phase diagram of CLRO. 
Main conclusions are as follows: 

\begin{itemize}
\item Non-metallic ground state with canted antiferromagnetic order is 
realized for $0.00 \leq x < 0.15$. 
The ground state is in a broad sense Mott insulator, but more precisely 
consistent with the Anderson localization state with Coulomb interaction. 
\item Filling control of Ca$_{2}$RuO$_{4}$ drastically changes 
the metallic ground state ($x > 0.10$) adjacent to CAF-NM ground state 
compared with purely bandwidth controlled CSRO. 
We did not find any sign of superconductivity at least down to 50 mK. 
FM fluctuation is strong in this region. 
\end{itemize}

\section*{Acknowledgements}
We would like to thank T. Ishiguro for his support in many aspects. 
We acknowledge useful discussion with S. Nakatsuji and his technical support. 
We appreciate T. Mizokawa for fruitful discussion. 
We are grateful to K. Kosuge, K. Yoshimura and M. Kato for permitting 
the authors to use EDX apparatus and for their kind help. 
One of the authors (H. F. ) is supported by JSPS Research Fellowships for 
Young Scientists.


\begin{thebibliography}{99}
\bibitem{mae1} Y. Maeno, H. Hashimoto, K. Yoshida, S. Nishizaki, 
T. Fujita, J. G. Bednortz and F. Lichtenberg: 
Nature {\bf 372} (1994) 532.
\bibitem{ish1} K. Ishida, H. Mukuda, Y. Kitaoka, K. Asayama, Z. Q. Mao, 
Y. Mori and Y. Maeno: 
Nature {\bf 396} (1998) 658.
\bibitem{mac1} A. P. Mackenzie, R. K. W. Haselwimmer, A. W. Tyler, 
G. G. Lonzarich, Y. Mori, S. Nishizaki and Y. Maeno: 
Phys. Rev. Lett. {\bf 80} (1998) 161.
\bibitem{mao1} Z. Q. Mao , Y. Maeno, S. NishiZaki, T. Akima and T. Ishiguro: 
Phys. Rev. Lett. {\bf 84} (2000) 991.
\bibitem{nis1} S. NishiZaki, Y. Maeno and Z. Q. Mao: 
J. Phys. Soc. Jpn. {\bf 69} (2000) 572.
\bibitem{mae2} Y. Maeno, K. Yoshida, H. Hashimoto, S. Nishizaki, S. Ikeda, 
M. Nohara, T. Fujita, A. P. Mackenzie, N. E. Hussey, J. G. Bednortz and 
F. Lichtenberg: 
J. Phys. Soc. Jpn. {\bf 66} (1997) 1405.
\bibitem{muk1} H. Mukuda, K. Ishida, Y. Kitaoka, K. Asayama, Z. Q. Mao, 
Y. Mori and Y. Maeno: 
J. Phys. Soc. Jpn. {\bf 67} (1998) 3945.
\bibitem{nak1} S. Nakatsuji, S. Ikeda and Y. Maeno: 
J. Phys. Soc. Jpn. {\bf 66} (1997) 1868.
\bibitem{cao1} G. Cao, S. McCall, M. Shepard, J. E. Crow and R. P. Guertin: 
Phys. Rev. B {\bf 56} (1997) R2916.
\bibitem{bra1} M. Braden, G. Andr$\rm{\acute{e}}$, S. Nakatsuji and Y. Maeno: 
Phys. Rev. B {\bf 58} (1998) 847. \label{refbra1}
\bibitem{ale1} C. S. Alexander, G. Cao, V. Dobrosavljevic, S. McCall, 
E. Lochner and R. P. Guertin: Phys. Rev. B {\bf 60} (1999) R8422.
\bibitem{fuk1} H. Fukazawa, S. Nakatsuji and Y. Maeno: 
Physica B {\bf 281 \& 282} (2000) 613.
\bibitem{nak2} S. Nakatsuji and Y. Maeno: 
Phys. Rev. Lett. {\bf 84} (2000) 2666.
\bibitem{nak3} S. Nakatsuji and Y. Maeno: 
to be published in Phys. Rev. B {\bf 62} (2000) Sep. 1 issue.
\bibitem{fri1} O. Friedt , M. Braden, G. Andr$\rm{\acute{e}}$, P. Adelmann, 
S. Nakatsuji and Y. Maeno: 
cond-mat/0007218.
\bibitem{fri2} The structural symmetry associated with longer lattice 
parameter $c$ along the long axis is denoted as L-Pbca: in contrast, 
the one with shorter $c$ parameter is denoted as S-Pbca.
\bibitem{ani1} V. I. Anisimov, I. A. Nekrasov, D. E. Kondakov, T. M. Rice and 
M. Sigrist: 
preprint.
\bibitem{cao2} G. Cao, S. McCall, V. Dobrosavljevic, C. S. Alexander, 
J. E. Crow and R. P. Guertin: 
Phys. Rev. B {\bf 61} (2000) R5053.
\bibitem{fzf1} The difference is especially significant 
in the bilayered perovskite Sr$_{3}$Ru$_{2}$O$_{7}$. 
For references, we cite two representative papers: 
for a FZ method, S. Ikeda, Y. Maeno, S. Nakatsuji, M. Kosaka and Y. Uwatoko: 
to be published in Phys. Rev. B {\bf 62} (2000) Sep. 1 issue. 
(cond-mat/0002147) ; 
for a flux method, G. Cao, S. McCall and J. E. Crow: 
Phys. Rev. B {\bf 55} (1997) R672. 
\bibitem{nic1} NEC Machinery, model SC-K15HD.
\bibitem{qua1} Quantum Design, model MPMS$_{5\rm{S}}$. 
We used an additional insert equipped with an oven for the measurement 
between 300 and 700 K.
\bibitem{mac2} A. P. Mackenzie, S. Ikeda, Y. Maeno, T. Fujita, S. R. Julian 
and G. G. Lonzarich: J. Phys. Soc. Jpn. {\bf 67} (1998) 385.
\bibitem{mor1} T. Moriya and A. Kawabata: 
J. Phys. Soc. Jpn. {\bf 35} (1973) 669.
\bibitem{maz1} I. I. Mazin and D. J. Singh: 
Phys. Rev. Lett. {\bf82} (1999) 4324.
\bibitem{maz2} I. I. Mazin and D. J. Singh: 
private communication.
\bibitem{miz1} T. Mizokawa, L. H. Tjeng, G. A. Sawatzky, G. Ghiringhelli, 
O. Tjernberg, N. B. Brookes, H. Fukazawa, S. Nakatsuji and Y. Maeno: 
preprint.
\bibitem{yos1} K. Yosida: \it{Theory of Magnetism}
\rm{(Springer-Verlag, Berlin, 1996)} pp. 38. 
\bibitem{fuk2} $\overline{\chi_{ab}}$ becomes smaller 
than $\chi_{c}$ again below about 150 K. 
However, it is owing to the reduction of $\chi_{[010]}$ associated 
with the evolution of short-range AF correlation 
above $\it{T}_{\rm{N}}$.~\cite{fuk1} 
\bibitem{mot1} N. F. Mott: 
\it{Metal-Insulator Transitions}\rm{, 2nd ed.} 
(Taylor \& Francis, London, 1990) Chap. 1.
\bibitem{efr1} A. L. Efros and B. I. Shklovskii: 
J. Phys. C {\bf 8} (1975) L49.
\bibitem{puc1} A. V. Puchkov, M. C. Schabel, D. N. Basov, T. Startseva, 
G. Cao, T. Timusk and Z. -X. Shen: 
Phys. Rev. Lett. {\bf 81} (1998) 2747.
\bibitem{ogu1} T. Oguchi: 
Phys. Rev. B {\bf 51} (1995) 1385.
\bibitem{nom1} T. Nomura and K. Yamada: 
J. Phys. Soc. Jpn {\bf 69} (2000) 1856.

\end{thebibliography}
\end{document}